\documentclass[prl,aps,superscriptaddress,showpacs,amssymb,
amsmath,twocolumn]{revtex4}
\usepackage{bm,graphicx}

\newcommand{\crete}{Department of Physics, University of Crete, 71003 
Heraklion, Greece}

\newcommand{\cern}{CERN, Physics Department, TH Division, CH-1211 Geneva 23, Switzerland}

\let\a=\alpha   \let\b=\beta   \let\g=\gamma   
         
        \let\m=\mu
\let\n=\nu                 
\let\s=\sigma        

\let\G=\Gamma  \let\D=\Delta   \let\L=\Lambda

\newcommand{\be}{\begin{equation}}
\newcommand{\ee}{\end{equation}}
\newcommand{\bea}{\begin{eqnarray}}
\newcommand{\eea}{\end{eqnarray}}
\newcommand{\ba}{\begin{array}}
\newcommand{\ea}{\end{array}}
\newcommand{\eq}[1]{Eq.~(\ref{#1})}
\newcommand{\fig}[1]{Fig.~\ref{#1}}

\def\tr{{\rm tr}}
\def\bn{\bar{n}}
\def\A5{(A_5)_{\rm lat}}
\begin{document}

\title{Non-perturbative mass spectrum of an extra-dimensional orbifold}

\author {N. \surname{Irges}}
\affiliation{\crete}

\author{F. \surname{Knechtli}}
\affiliation{\cern}

\begin{abstract}

We analyse non-perturbatively a five-dimensional $SU(2)$ gauge theory 
compactified
on the $S^1/\mathbb{Z}_2$ orbifold.
In particular, we present simulation results for the mass spectrum of 
the theory,
which contains a Higgs and a photon.
The Higgs mass is found to be free of divergences without fine-tuning.
The photon mass is non-zero, thus providing us with the
first lattice evidence for a Higgs mechanism derived from an extra 
dimension.
Data from the static potential are consistent with dimensional reduction at
low energies.

\end{abstract}


\pacs{11.10.Kk, 
      11.15.Ex, 
      11.15.Ha  
      \hfill CERN-PH-TH/2006-057
     }

\maketitle

\section{\label{s_intro} Introduction}

An attempt to embed the Standard Model in a more general
theory reveals subtleties associated with its Higgs sector such
as the fine-tuning problem and,
if for example the larger theory includes also gravity, the hierarchy 
problem. 
The former is essentially a reflection of the quadratic sensitivity
of the Higgs mass to the ultra-violet cut-off and
the latter refers to the mystery associated with the smallness of the
ratio of the electroweak and Planck scales $M_{\rm EW}/M_{\rm Pl}$. 
Supersymmetry provides a possible solution to the fine-tuning problem 
but at the
cost of introducing many new couplings and degrees of freedom into
the Standard Model. This is of course not necessarily a disaster, especially
if low-energy supersymmetry is confirmed at the LHC.
Since however the latter is not guaranteed, it is perhaps wise
to think of alternative scenarios.

In this Letter, we present results from the investigation of
a simple model, which gives
a possible explanation of the origin of the Higgs field
and at the same time does not suffer from a fine-tuning problem.
Since we carry out
our analysis in the context of gauge theories, we will not have
anything to say about the hierarchy problem.
Also, in order to illustrate in the simplest possible way
the underlying physics, we would like to postpone
technical details to a later work.

The model we will consider is
a five-dimensional pure $SU(2)$ gauge theory
with its fifth dimension compactified on the $S^1/\mathbb{Z}_2$ orbifold
\footnote{
The first use of orbifold geometries in physics is in
\cite{Dixon:1985jw}.}, and
with the $\mathbb{Z}_2$ acting as a reflection of the extra-dimensional 
coordinate.
It is possible to embed the $\mathbb{Z}_2$ action
into the gauge group so that it breaks on the orbifold boundaries
to a $U(1)$ subgroup, which results in the appearance of a
complex scalar field with the four-dimensional quantum numbers of
a Higgs field \cite{Antoniadis:2001cv}.
At the classical level the scalar is massless.
However at 1 loop, a dynamically generated potential is formed
and the scalar can in principle further break the gauge group spontaneously
by taking a vacuum expectation value.
Perturbative studies have revealed 
that the presence of bulk fermions or scalars  is
necessary for this mechanism to work
\cite{Hosotani:1983xw,Hosotani:1988bm,Kubo:2001zc} but a non-zero
Higgs mass is generated anyway, just as one would expect by
trivially extending results obtained in
finite-temperature field theory \cite{ZinnJustin-Book} or as one can 
verify by a computation in the Kaluza--Klein framework
\cite{vonGersdorff:2002as,Cheng:2002iz}. 
In fact, the Kaluza--Klein expansion, being a gauge
in which the states in the Hilbert space are diagonalized
with respect to their four-dimensional quantum numbers
is the one that best fits the perturbative
approach to compactified extra-dimensional field theories. 
In a non-perturbative approach it seems necessary, though, to keep
the entire gauge invariance intact, and thus the Kaluza--Klein
construction is less useful.

\section{\label{s_lat} Orbifold on the lattice}

As an alternative approach to perturbation theory,
we use the non-perturbative definition of
five-dimensional $SU(N)$ gauge theories compactified
on the $S^1/\mathbb{Z}_2$ orbifold \cite{Irges:2004gy} and analyse the
system via lattice simulations
\footnote{Earlier lattice investigations
of five-dimensional systems of geometries (including anisotropic, 
layered, warped)
without boundaries can be found in
\cite{Creutz:1979dw,Beard:1997ic,Ejiri:2000fc,Farakos:2002zb,Fu:1983ei,Dimopoulos:2000ej}.
The localization of gauge fields in the presence of a warped extra dimension 
\cite{Laine:2002rh}
or a domain wall in the extra dimension \cite{Laine:2004ji} has been 
investigated on
the lattice in $2+1$ dimensions.
}. The first signal of interesting non-perturbative physics can be
anticipated by looking at the lattice coupling
\be
\b = \frac{2N}{g_5^2} a,\label{beta}
\ee
where $a$ is the lattice spacing
(which provides the inverse cut-off $1/\L$) and
$g_5$ is the five-dimensional gauge coupling, which has mass dimension 
$-1/2$.
The latter can be thought of as an effective coupling at the cut-off scale.
Naive dimensional analysis tells us
that as $\b$ decreases with $g_5$ fixed, the lattice spacing
also decreases and the dimensionless bare coupling $g_0=g_5 \sqrt{1/a}$ 
blows up.
One would therefore expect
to find the perturbative regime in the large-$\b$ region
where the lattice spacing is large and the bare coupling small.
The compactification scale is $1/R$, with $R$ the radius of $S^1$ and
a separation from the cut-off scale requires $a/R\ll1$.
Increasing $\b$ would require an increase
also of $R$, which drives the fifth dimension to its
decompactification limit; as a result
the system degenerates to a theory of massless photons.
A general lesson from the above discussion then
is that moving towards
the perturbative regime is expected to enhance the cut-off
effects (appearing as $E/\L$ at low energies $E$ in the sense
of an effective action) and decompactify the theory,
whereas moving in the opposite direction, i.e. towards small $\b$,
is expected to suppress the
cut-off effects and drive the system into a compactified but 
non-perturbative regime.
Eventually a phase transition is reached at a critical value of $\b$
\cite{Creutz:1979dw,Beard:1997ic,Knechtli:2005dw}, where the cut-off reaches
its maximal value.
The viability of perturbative extra-dimensional extensions of the 
Standard Model
then essentially relies on the existence of an overlap between these two 
regimes
and clearly a computational method that can probe both of them, such as 
the lattice,
could provide us with a unique insight.

Gauge theories on the orbifold can be discretized on the lattice
\cite{Irges:2004gy,Knechtli:2005dw}.
One starts with a gauge theory formulated on a five-dimensional torus with
lattice spacing $a$ and periodic boundary conditions in all directions 
$M=0,1,2,3,5$.
The spatial directions ($M=1,2,3$) have length $L$,
the time-like direction ($M=0$) has length $T$, and the
extra dimension ($M=5$) has length $2\pi R$. The coordinates of the 
points are
labelled by integers $n\equiv\{n_M\}$ and the gauge field is the set of link
variables $\{U(n,M)\in SU(N)\}$.
The latter are related to a gauge potential 
$A_M$ in the Lie algebra of $SU(N)$
by $U(n,M) = \exp\{a A_M(n)\}$.
Embedding the orbifold action in the gauge field
on the lattice amounts to imposing on the links the $\mathbb{Z}_2$ 
projection
\be
(1-\G)U(n,M) = 0 \,, \label{orbiproj}
\ee
where $\G = {\cal R} {\cal T}_g$. Here, ${\cal R}$ is the
reflection operator that acts as ${\cal 
R}\,n=(n_\mu,-n_5)\equiv\bn\;(\mu=0,1,2,3)$
on the lattice and as ${\cal R}\,U(n,\m) = U (\bn,\m)$
and ${\cal R}\,U(n,5) = U^\dagger(\bn-{\hat 5},5)$ on the links.
The group conjugation ${\cal T}_g$ acts only on the links, as
${\cal T}_g U(n,M) = g U(n,M) g^{-1}$,
where $g$ is a constant $SU(N)$ matrix with the property that
$g^2$ is an element of the centre of $SU(N)$.
For $SU(2)$ we will take $g=-i\s^3$. Only gauge transformations 
$\{\Omega(n)\}$
satisfying $(1-\G)\Omega=0$ are consistent with \eq{orbiproj}. This 
means that
at the orbifold fixed points, for which $n_5=0$ or $n_5=\pi R/a=N_5$, the
gauge group is broken to the subgroup that commutes with $g$. For 
$SU(2)$ this
is the $U(1)$ subgroup parametrized by $\exp(i\phi\s^3)$, where $\phi$ 
are compact
phases.

After the projection in \eq{orbiproj},
the fundamental domain is the strip
$I_0 = \{n_\m , 0\le n_5 \le N_5\}$.
The gauge-field action on $I_0$ is taken to be the Wilson action
\be
S_W^{\rm orb}[U] = \frac{\b}{2N} \sum_p w(p)\; {\tr}\, \{1-U(p)\}, 
\label{wila}
\ee
where the sum runs over all oriented plaquettes $U(p)$ in $I_0$.
The weight $w(p)$ is 1/2 if $p$ is a plaquette in the $(\m\n)$ planes
at $n_5=0$ and $n_5=N_5$, and 1 in all other cases.
At the orbifold boundary planes
Dirichlet boundary conditions are imposed on the gauge links
\bea
 U(n,\mu) & = & g\,U(n,\mu)\,g^{-1} \,. \label{Dbclat}
\eea
The gauge variables at the boundaries are not fixed but are
restricted to the subgroup of $SU(N)$, invariant under ${\cal T}_g$.
The Wilson action together with these boundary conditions reproduce the 
correct
naive continuum gauge action and boundary conditions on the components 
of the
five-dimensional gauge potential \cite{Irges:2004gy}.
For example, for $SU(2)$, $A_\m^{3}$ (``photon'')
and $A_5^{1,2}$ (``Higgs'') satisfy Neumann boundary conditions and
$A_\m^{1,2}$ and $A_5^{3}$ Dirichlet ones.

\section{Lattice operators \label{s_op}}

If the fifth dimension were infinite, the gauge links $U(n,5)$ would be
gauge-equivalent to the identity, which corresponds to the continuum
axial gauge $A_5\equiv0$.
On the circle $S^1$ one can gauge-transform
$U(n,5)$ to an $n_5$-independent matrix $V(n_\mu)$ that
satisfies $L=V^{2N_5}$, where $L(n_\mu)$
is the Polyakov line winding around the extra dimension.
Therefore an extra-dimensional potential
$\A5$ can be defined on the lattice, through $V=\exp\{a\A5\}$, as
\be
 a\A5 = \frac{1}{4N_5}(L-L^\dagger) + {\rm O}(a^3) \,. \label{A5lat}
\ee
At finite lattice spacing the O($a^3$) corrections in \eq{A5lat}
are neglected. By imposing the orbifold projection \eq{orbiproj}
on the links building $L$, it is straightforward to obtain a definition
of $\A5$ on the $S^1/\mathbb{Z}_2$ orbifold. For
the adjoint index of $\A5$ to be separated into even and odd components under the
conjugation ${\cal T}_g$, the Polyakov line must start and end at one
of the boundaries. The odd components under ${\cal T}_g$ represent the
Higgs field
\be
 \Phi = [\A5 , g] \,, \label{Higgsfield}
\ee
which has the same gauge transformation as a field strength tensor.
A gauge-invariant operator for the Higgs field, which can be used to
extract its mass, is $\tr\{\Phi\Phi^\dagger\}$.

Five-dimensional gauge invariance strictly forbids a
boundary mass counterterm in the action
\cite{vonGersdorff:2002us,vonGersdorff:2002rg,Irges:2004gy}.
Notice that if a boundary mass term was allowed then an additional
mass parameter $\m$ would have to appear in the lattice action
through an explicit boundary $\m^2/a^2$ term.
Changing $\m$ would have to be done in a fine-tuned way
in order to keep the physical Higgs mass $m_h$ constant.
This is the lattice version of the Higgs fine-tuning problem.
The contribution to the mass of the Higgs particle(s)
is therefore expected to come from bulk and bulk--boundary effects,
which is reflected by the non-locality of
the operator $\Phi$ in \eq{Higgsfield}.

The 1-loop Higgs effective potential in the pure gauge theory
does not lead to the spontaneous symmetry breaking of the remnant
gauge group \cite{Kubo:2001zc}. In this case
the Higgs mass is given by (for general $N$) \cite{vonGersdorff:2002as}
\be
m_h R = \frac{c}{\sqrt{N_5 \b}},\label{lattHiggs}
\ee
where $c=3/(4\pi^2) \sqrt{N\zeta(3)C_2(N)}$ and $C_2(N)=(N^2-1)/(2N)$.
In the same spirit, the first excited (Kaluza--Klein)
state in this sector is expected to appear
split from the ground state by $(\D m)\, a = \pi /N_5$,
the second with a mass splitting twice that, and so forth.
Since the five-dimensional theory is non-renormalizable
it is non-trivial if \eq{lattHiggs}
remains cut-off-insensitive at higher orders in perturbation theory
\footnote{This is not a statement that one can easily find
in the literature. From our point of view,
apart from certain quite convincing arguments
regarding pure bulk effects \cite{Scrucca:2003ra},
there is no rigorous proof of the finiteness of the Higgs mass
when bulk--boundary mixing effects are included. Indeed the latter produces
at 2 loops a logarithmic sensitivity to the cut-off 
\cite{vonGersdorff:2005ce}.
}.
Also it is not clear whether the absence of spontaneous symmetry breaking
at 1 loop pertains at higher orders or in fact non-perturbatively.

A quantity that could settle this last question is the mass of the photon.
The lattice photon field can be constructed from the Higgs field in
analogy to the standard four-dimensional Higgs model \cite{Montvay:1984wy}.
We define the $SU(2)$-valued quantity
$\a = \Phi/\sqrt{\det(\Phi)}$
and from it the gauge-invariant field
\be
W_k = \tr\{g V_k\} \,,\quad k=1,2,3\,,\label{lattphoton}
\ee
where $V_k(n)=U^\dagger(n,k) \a^\dagger(n+{\hat k}) U(n,k) \a(n)$ is
evaluated at one of the boundaries. $W_k$ is even under ${\cal T}_g$
and it is clear that it has the correct quantum numbers
to be identified as the lattice operator that
corresponds to the continuum $\mathbb{Z}_2$ even $U(1)$ gauge-field
component $A_k^3$.

We build variational bases of Higgs and photon operators
with the help of smeared gauge links and
alternative definitions of (smeared) Higgs fields.
In each basis
the masses are extracted from connected time-correlation matrices,
using the technique of \cite{Luscher:1990ck}.

The static potential can be extracted from four-dimensional Wilson loops
in the slices orthogonal to the extra dimension.
These are operators sensitive to the confinement/deconfinement properties
of the system, its dimensionality and spontaneous symmetry breaking.
For example if the system is in a deconfined, dimensionally reduced and
spontaneously broken phase one would expect to see a four-dimensional Yukawa
static potential.

\section{The mass spectrum \label{s_spec}}

\begin{figure}
\includegraphics[width=3in]{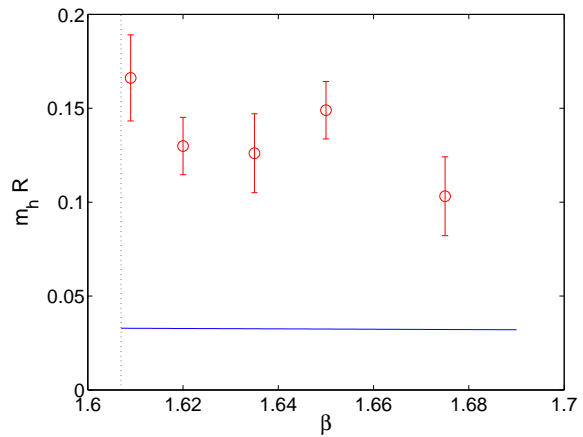}
\caption{\label{f_mh} The Higgs mass.}
\end{figure}

\begin{figure}
\includegraphics[width=3in]{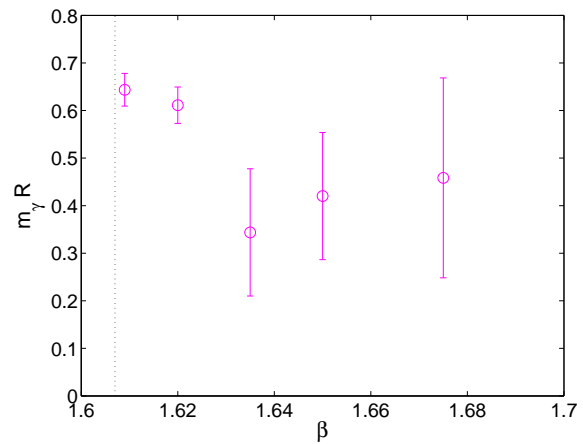}
\caption{\label{f_mpho} The photon mass.}
\end{figure}

In this section we present results from simulations of the $SU(2)$
theory and, specifically, we compute the masses of the Higgs and photon.
The free parameters of the model are essentially $\beta$ and
$N_5$. We choose the lattice sizes to be $T/a=64$,
$L/a=16$, and $N_5=6$. The algorithm uses heatbath and overrelaxation
updates for $SU(2)$ bulk and $U(1)$ boundary links.
Simulations are performed in the deconfined
phase (large $\beta$) approaching the phase transition, which is
located at $\beta_c=1.607$ (marked by a vertical dotted line in
\fig{f_mh} and \fig{f_mpho}). In the confined phase the signal for the
effective masses of the particles is lost.
The statistics is between 6000 and 10000 measurements separated by
1 heatbath and 8 overrelaxation sweeps.

Figure \ref{f_mh} shows the Higgs mass $m_h$ in units of $1/R$
as a function of $\beta$.
The solid horizontal line corresponds to the 1-loop formula \eq{lattHiggs}.
The first observation we would like to make
is that the Higgs mass can be measured
without any fine-tuning of the lattice parameters. The second is that it
does not diverge as the cut-off is increased by approaching the phase
transition. The third observation is that the Higgs mass in units of $1/R$
decreases slowly
as we move away from
the phase transition.

Figure \ref{f_mpho} shows the photon mass $m_\g$ in units of $1/R$
as a function of $\beta$.
Contrary to the 1-loop prediction, it is non-zero, indicating spontaneous
symmetry breaking in the pure gauge theory. This is the first 
non-perturbative
evidence for the Higgs mechanism originating from an extra dimension.
The photon mass in units of $1/R$ is constant close to the phase transition
and decreases at larger $\b$, where it becomes more difficult to extract.
In this simple model the photon mass is larger than the Higgs mass.
It would be interesting to see if this is a generic property since
in phenomenological applications one would like the
Higgs to be heavier than the vector bosons.

\section{The static potential \label{s_pot}}

In this section we discuss the results for the static potential in the
four-dimensional slices. Simulations were performed on lattices of sizes
$T/a=32$, $L/a=16$, and $N_5=6$.
The statistics is 4000 measurements.
\begin{figure}
\includegraphics[width=3in]{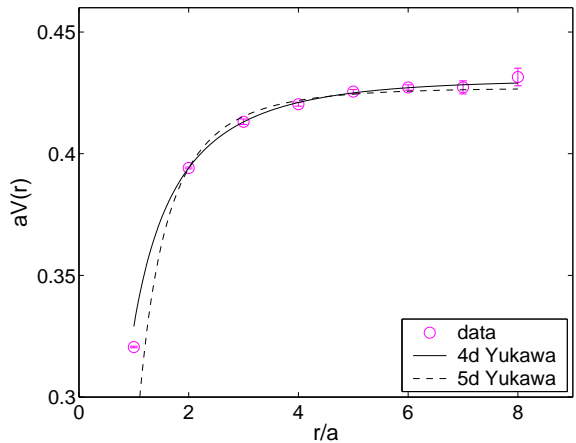}
\caption{\label{f_pot} U(1) static potential at the boundary.}
\end{figure}

Figure \ref{f_pot} shows the static potential on the boundary slice $n_5=0$ for 
$\beta=1.609$.
Since the photon mass is non-zero, fits to four- ($-c_1\exp(-m_\gamma 
r)/r+c_0$, solid line)
and five- ($-d_1K_1(m_\gamma r)/r+d_0$, where $K_1$ is a modified Bessel function, dashed line) dimensional 
Yukawa potentials
are performed, using $m_\gamma R=0.646$ from \fig{f_mpho}.
The point at $r/a=1$ is neglected.
The data are consistent with spontaneous symmetry breaking and
the minimum $\chi^2/{\rm dof}=0.35$
is obtained for the
four-dimensional Yukawa potential. For the five-dimensional Yukawa potential
we get $\chi^2/n_{\rm df}=2.6$.
Thus the data favour dimensional reduction.
Ignoring spontaneous symmetry breaking, acceptable fits to the data 
are also obtained with a five-dimensional ($-e_1/r^2+e_0$) or
a four-dimensional ($-f_1/r+f_0$) Coulomb form.

The potential in the four-dimensional slices in the bulk has larger
errors and can be fitted equally well to any of the potential forms 
mentioned above.


\begin{acknowledgments}
We thank B. Bunk for his help in the construction of the programming code.
We are grateful to M. L\"uscher for discussions and helpful suggestions.
We thank the Swiss National Supercomputing Centre (CSCS) in Manno 
(Switzerland)
for allocating computer resources to this project.
N. Irges thanks CERN for hospitality.
\end{acknowledgments}


\bibliography{Higgs}

\begin{thebibliography}{24}
\expandafter\ifx\csname natexlab\endcsname\relax\def\natexlab#1{#1}\fi
\expandafter\ifx\csname bibnamefont\endcsname\relax
  \def\bibnamefont#1{#1}\fi
\expandafter\ifx\csname bibfnamefont\endcsname\relax
  \def\bibfnamefont#1{#1}\fi
\expandafter\ifx\csname citenamefont\endcsname\relax
  \def\citenamefont#1{#1}\fi
\expandafter\ifx\csname url\endcsname\relax
  \def\url#1{\texttt{#1}}\fi
\expandafter\ifx\csname urlprefix\endcsname\relax\def\urlprefix{URL }\fi
\providecommand{\bibinfo}[2]{#2}
\providecommand{\eprint}[2][]{\url{#2}}

\bibitem[{\citenamefont{Antoniadis et~al.}(2001)\citenamefont{Antoniadis,
  Benakli, and Quiros}}]{Antoniadis:2001cv}
\bibinfo{author}{\bibfnamefont{I.}~\bibnamefont{Antoniadis}},
  \bibinfo{author}{\bibfnamefont{K.}~\bibnamefont{Benakli}}, \bibnamefont{and}
  \bibinfo{author}{\bibfnamefont{M.}~\bibnamefont{Quiros}},
  \bibinfo{journal}{New J. Phys.} \textbf{\bibinfo{volume}{3}},
  \bibinfo{pages}{20} (\bibinfo{year}{2001}), \eprint{hep-th/0108005}.

\bibitem[{\citenamefont{Hosotani}(1983)}]{Hosotani:1983xw}
\bibinfo{author}{\bibfnamefont{Y.}~\bibnamefont{Hosotani}},
  \bibinfo{journal}{Phys. Lett.} \textbf{\bibinfo{volume}{B126}},
  \bibinfo{pages}{309} (\bibinfo{year}{1983}).

\bibitem[{\citenamefont{Hosotani}(1989)}]{Hosotani:1988bm}
\bibinfo{author}{\bibfnamefont{Y.}~\bibnamefont{Hosotani}},
  \bibinfo{journal}{Ann. Phys.} \textbf{\bibinfo{volume}{190}},
  \bibinfo{pages}{233} (\bibinfo{year}{1989}).

\bibitem[{\citenamefont{Kubo et~al.}(2002)\citenamefont{Kubo, Lim, and
  Yamashita}}]{Kubo:2001zc}
\bibinfo{author}{\bibfnamefont{M.}~\bibnamefont{Kubo}},
  \bibinfo{author}{\bibfnamefont{C.~S.} \bibnamefont{Lim}}, \bibnamefont{and}
  \bibinfo{author}{\bibfnamefont{H.}~\bibnamefont{Yamashita}},
  \bibinfo{journal}{Mod. Phys. Lett.} \textbf{\bibinfo{volume}{A17}},
  \bibinfo{pages}{2249} (\bibinfo{year}{2002}), \eprint{hep-ph/0111327}.

\bibitem[{\citenamefont{Zinn-Justin}(2002)}]{ZinnJustin-Book}
\bibinfo{author}{\bibfnamefont{J.}~\bibnamefont{Zinn-Justin}},
  \emph{\bibinfo{title}{Quantum Field Theory and Critical Phenomena}}
  (\bibinfo{publisher}{International Series of Monographs on Physics --- Vol.
  113, Clarendon Press}, \bibinfo{address}{Oxford}, \bibinfo{year}{2002}),
  \bibinfo{edition}{4th} ed., ISBN \bibinfo{isbn}{0-19-850923-5}.

\bibitem[{\citenamefont{von Gersdorff
  et~al.}(2002{\natexlab{a}})\citenamefont{von Gersdorff, Irges, and
  Quiros}}]{vonGersdorff:2002as}
\bibinfo{author}{\bibfnamefont{G.}~\bibnamefont{von Gersdorff}},
  \bibinfo{author}{\bibfnamefont{N.}~\bibnamefont{Irges}}, \bibnamefont{and}
  \bibinfo{author}{\bibfnamefont{M.}~\bibnamefont{Quiros}},
  \bibinfo{journal}{Nucl. Phys.} \textbf{\bibinfo{volume}{B635}},
  \bibinfo{pages}{127} (\bibinfo{year}{2002}{\natexlab{a}}),
  \eprint{hep-th/0204223}.

\bibitem[{\citenamefont{Cheng et~al.}(2002)\citenamefont{Cheng, Matchev, and
  Schmaltz}}]{Cheng:2002iz}
\bibinfo{author}{\bibfnamefont{H.-C.} \bibnamefont{Cheng}},
  \bibinfo{author}{\bibfnamefont{K.~T.} \bibnamefont{Matchev}},
  \bibnamefont{and} \bibinfo{author}{\bibfnamefont{M.}~\bibnamefont{Schmaltz}},
  \bibinfo{journal}{Phys. Rev.} \textbf{\bibinfo{volume}{D66}},
  \bibinfo{pages}{036005} (\bibinfo{year}{2002}), \eprint{hep-ph/0204342}.

\bibitem[{\citenamefont{Irges and Knechtli}(2005)}]{Irges:2004gy}
\bibinfo{author}{\bibfnamefont{N.}~\bibnamefont{Irges}} \bibnamefont{and}
  \bibinfo{author}{\bibfnamefont{F.}~\bibnamefont{Knechtli}},
  \bibinfo{journal}{Nucl. Phys.} \textbf{\bibinfo{volume}{B719}},
  \bibinfo{pages}{121} (\bibinfo{year}{2005}), \eprint{hep-lat/0411018}.

\bibitem[{\citenamefont{Creutz}(1979)}]{Creutz:1979dw}
\bibinfo{author}{\bibfnamefont{M.}~\bibnamefont{Creutz}},
  \bibinfo{journal}{Phys. Rev. Lett.} \textbf{\bibinfo{volume}{43}},
  \bibinfo{pages}{553} (\bibinfo{year}{1979}).

\bibitem[{\citenamefont{Beard et~al.}(1998)}]{Beard:1997ic}
\bibinfo{author}{\bibfnamefont{B.~B.} \bibnamefont{Beard}}
  \bibnamefont{et~al.}, \bibinfo{journal}{Nucl. Phys. Proc. Suppl.}
  \textbf{\bibinfo{volume}{63}}, \bibinfo{pages}{775} (\bibinfo{year}{1998}),
  \eprint{hep-lat/9709120}.

\bibitem[{\citenamefont{Knechtli et~al.}(2005)\citenamefont{Knechtli, Bunk, and
  Irges}}]{Knechtli:2005dw}
\bibinfo{author}{\bibfnamefont{F.}~\bibnamefont{Knechtli}},
  \bibinfo{author}{\bibfnamefont{B.}~\bibnamefont{Bunk}}, \bibnamefont{and}
  \bibinfo{author}{\bibfnamefont{N.}~\bibnamefont{Irges}},
  \bibinfo{journal}{PoS} \textbf{\bibinfo{volume}{LAT2005}},
  \bibinfo{pages}{280} (\bibinfo{year}{2005}), \eprint{hep-lat/0509071}.

\bibitem[{\citenamefont{von Gersdorff et~al.}(2003)\citenamefont{von Gersdorff,
  Irges, and Quiros}}]{vonGersdorff:2002us}
\bibinfo{author}{\bibfnamefont{G.}~\bibnamefont{von Gersdorff}},
  \bibinfo{author}{\bibfnamefont{N.}~\bibnamefont{Irges}}, \bibnamefont{and}
  \bibinfo{author}{\bibfnamefont{M.}~\bibnamefont{Quiros}},
  \bibinfo{journal}{Phys. Lett.} \textbf{\bibinfo{volume}{B551}},
  \bibinfo{pages}{351} (\bibinfo{year}{2003}), \eprint{hep-ph/0210134}.

\bibitem[{\citenamefont{von Gersdorff
  et~al.}(2002{\natexlab{b}})\citenamefont{von Gersdorff, Irges, and
  Quiros}}]{vonGersdorff:2002rg}
\bibinfo{author}{\bibfnamefont{G.}~\bibnamefont{von Gersdorff}},
  \bibinfo{author}{\bibfnamefont{N.}~\bibnamefont{Irges}}, \bibnamefont{and}
  \bibinfo{author}{\bibfnamefont{M.}~\bibnamefont{Quiros}}
  (\bibinfo{year}{2002}{\natexlab{b}}), \eprint{hep-ph/0206029}.

\bibitem[{\citenamefont{Montvay}(1985)}]{Montvay:1984wy}
\bibinfo{author}{\bibfnamefont{I.}~\bibnamefont{Montvay}},
  \bibinfo{journal}{Phys. Lett.} \textbf{\bibinfo{volume}{B150}},
  \bibinfo{pages}{441} (\bibinfo{year}{1985}).

\bibitem[{\citenamefont{L{\"u}scher and Wolff}(1990)}]{Luscher:1990ck}
\bibinfo{author}{\bibfnamefont{M.}~\bibnamefont{L{\"u}scher}} \bibnamefont{and}
  \bibinfo{author}{\bibfnamefont{U.}~\bibnamefont{Wolff}},
  \bibinfo{journal}{Nucl. Phys.} \textbf{\bibinfo{volume}{B339}},
  \bibinfo{pages}{222} (\bibinfo{year}{1990}).

\bibitem[{\citenamefont{Dixon et~al.}(1985)\citenamefont{Dixon, Harvey, Vafa,
  and Witten}}]{Dixon:1985jw}
\bibinfo{author}{\bibfnamefont{L.~J.} \bibnamefont{Dixon}},
  \bibinfo{author}{\bibfnamefont{J.~A.} \bibnamefont{Harvey}},
  \bibinfo{author}{\bibfnamefont{C.}~\bibnamefont{Vafa}}, \bibnamefont{and}
  \bibinfo{author}{\bibfnamefont{E.}~\bibnamefont{Witten}},
  \bibinfo{journal}{Nucl. Phys.} \textbf{\bibinfo{volume}{B261}},
  \bibinfo{pages}{678} (\bibinfo{year}{1985}).

\bibitem[{\citenamefont{Ejiri et~al.}(2000)\citenamefont{Ejiri, Kubo, and
  Murata}}]{Ejiri:2000fc}
\bibinfo{author}{\bibfnamefont{S.}~\bibnamefont{Ejiri}},
  \bibinfo{author}{\bibfnamefont{J.}~\bibnamefont{Kubo}}, \bibnamefont{and}
  \bibinfo{author}{\bibfnamefont{M.}~\bibnamefont{Murata}},
  \bibinfo{journal}{Phys. Rev.} \textbf{\bibinfo{volume}{D62}},
  \bibinfo{pages}{105025} (\bibinfo{year}{2000}), \eprint{hep-ph/0006217}.

\bibitem[{\citenamefont{Farakos et~al.}(2003)\citenamefont{Farakos,
  de~Forcrand, Korthals~Altes, Laine, and Vettorazzo}}]{Farakos:2002zb}
\bibinfo{author}{\bibfnamefont{K.}~\bibnamefont{Farakos}},
  \bibinfo{author}{\bibfnamefont{P.}~\bibnamefont{de~Forcrand}},
  \bibinfo{author}{\bibfnamefont{C.~P.} \bibnamefont{Korthals~Altes}},
  \bibinfo{author}{\bibfnamefont{M.}~\bibnamefont{Laine}}, \bibnamefont{and}
  \bibinfo{author}{\bibfnamefont{M.}~\bibnamefont{Vettorazzo}},
  \bibinfo{journal}{Nucl. Phys.} \textbf{\bibinfo{volume}{B655}},
  \bibinfo{pages}{170} (\bibinfo{year}{2003}), \eprint{hep-ph/0207343}.

\bibitem[{\citenamefont{Fu and Nielsen}(1984)}]{Fu:1983ei}
\bibinfo{author}{\bibfnamefont{Y.~K.} \bibnamefont{Fu}} \bibnamefont{and}
  \bibinfo{author}{\bibfnamefont{H.~B.} \bibnamefont{Nielsen}},
  \bibinfo{journal}{Nucl. Phys.} \textbf{\bibinfo{volume}{B236}},
  \bibinfo{pages}{167} (\bibinfo{year}{1984}).

\bibitem[{\citenamefont{Dimopoulos et~al.}(2001)\citenamefont{Dimopoulos,
  Farakos, Kehagias, and Koutsoumbas}}]{Dimopoulos:2000ej}
\bibinfo{author}{\bibfnamefont{P.}~\bibnamefont{Dimopoulos}},
  \bibinfo{author}{\bibfnamefont{K.}~\bibnamefont{Farakos}},
  \bibinfo{author}{\bibfnamefont{A.}~\bibnamefont{Kehagias}}, \bibnamefont{and}
  \bibinfo{author}{\bibfnamefont{G.}~\bibnamefont{Koutsoumbas}},
  \bibinfo{journal}{Nucl. Phys.} \textbf{\bibinfo{volume}{B617}},
  \bibinfo{pages}{237} (\bibinfo{year}{2001}), \eprint{hep-th/0007079}.

\bibitem[{\citenamefont{Laine et~al.}(2003)\citenamefont{Laine, Meyer,
  Rummukainen, and Shaposhnikov}}]{Laine:2002rh}
\bibinfo{author}{\bibfnamefont{M.}~\bibnamefont{Laine}},
  \bibinfo{author}{\bibfnamefont{H.~B.} \bibnamefont{Meyer}},
  \bibinfo{author}{\bibfnamefont{K.}~\bibnamefont{Rummukainen}},
  \bibnamefont{and}
  \bibinfo{author}{\bibfnamefont{M.}~\bibnamefont{Shaposhnikov}},
  \bibinfo{journal}{JHEP} \textbf{\bibinfo{volume}{01}}, \bibinfo{pages}{068}
  (\bibinfo{year}{2003}), \eprint{hep-ph/0211149}.

\bibitem[{\citenamefont{Laine et~al.}(2004)\citenamefont{Laine, Meyer,
  Rummukainen, and Shaposhnikov}}]{Laine:2004ji}
\bibinfo{author}{\bibfnamefont{M.}~\bibnamefont{Laine}},
  \bibinfo{author}{\bibfnamefont{H.~B.} \bibnamefont{Meyer}},
  \bibinfo{author}{\bibfnamefont{K.}~\bibnamefont{Rummukainen}},
  \bibnamefont{and}
  \bibinfo{author}{\bibfnamefont{M.}~\bibnamefont{Shaposhnikov}},
  \bibinfo{journal}{JHEP} \textbf{\bibinfo{volume}{04}}, \bibinfo{pages}{027}
  (\bibinfo{year}{2004}), \eprint{hep-ph/0404058}.

\bibitem[{\citenamefont{Scrucca et~al.}(2003)\citenamefont{Scrucca, Serone, and
  Silvestrini}}]{Scrucca:2003ra}
\bibinfo{author}{\bibfnamefont{C.~A.} \bibnamefont{Scrucca}},
  \bibinfo{author}{\bibfnamefont{M.}~\bibnamefont{Serone}}, \bibnamefont{and}
  \bibinfo{author}{\bibfnamefont{L.}~\bibnamefont{Silvestrini}},
  \bibinfo{journal}{Nucl. Phys.} \textbf{\bibinfo{volume}{B669}},
  \bibinfo{pages}{128} (\bibinfo{year}{2003}), \eprint{hep-ph/0304220}.

\bibitem[{\citenamefont{von Gersdorff and
  Hebecker}(2005)}]{vonGersdorff:2005ce}
\bibinfo{author}{\bibfnamefont{G.}~\bibnamefont{von Gersdorff}}
  \bibnamefont{and} \bibinfo{author}{\bibfnamefont{A.}~\bibnamefont{Hebecker}},
  \bibinfo{journal}{Nucl. Phys.} \textbf{\bibinfo{volume}{B720}},
  \bibinfo{pages}{211} (\bibinfo{year}{2005}), \eprint{hep-th/0504002}.

\end{thebibliography}


\end{document}